\documentclass[aps,prb,reprint,twocolumn,showpacs,floatfix,superscriptaddress,nofootinbib]{revtex4}
\usepackage{graphicx}
\usepackage{amssymb}
\usepackage{amsmath}
\usepackage{lmodern}
\usepackage{color}
\usepackage{hyperref}
\usepackage{empheq}
\usepackage{epstopdf}
\usepackage{amsmath}

\begin{document}

\title{Josephson transmission line revisited}

\author{Eugene Kogan}
\email{Eugene.Kogan@biu.ac.il}
\affiliation{Department of Physics, Bar-Ilan University, Ramat-Gan 52900, Israel}
\affiliation{Max-Planck-Institut fur Physik komplexer Systeme
Dresden 01187, Germany}

\begin{abstract}
We consider the series-connected  Josephson transmission line (JTL), constructed from  Josephson junctions,  capacitors and (possibly) resistors. We calculate the velocity of shocks   in the discrete lossy JTL. We study thoroughly the continuum and the quasi-continuum approximations to the discrete JTL, both lossless and lossy. In the framework of these approximations we show that the compact travelling waves  in the lossless JTL are the kinks and the solitons, and calculate their velocities. On top of each of the above mentioned approximations, we propose
the simple wave approximation, which   decouples  the JTL equations   into two separate equations for the right- and left-going waves. The approximation, in particular, allows to easily consider the formation of shocks in the lossy JTL.

\end{abstract}

\date{\today}

\maketitle

\section{Introduction}
\label{introd}

The concept that in a nonlinear wave propagation system
the various parts of the wave travel with different
velocities, and that wave fronts (or tails) can sharpen
into shock waves, is deeply imbedded in the classical
theory of fluid dynamics \cite{whitham}.
The methods developed in that field can be profitably used
to study signal propagation in nonlinear transmission lines
\cite{french,nouri,neto,nikoo,silva,wang,rangel,kyuregyan,akem,fairbanks}.
In the early studies of shock waves in  transmission lines, the
origin of the nonlinearity was due to nonlinear capacitance
in the circuit \cite{landauer,peng,rabinovich}.

Interesting and potentially important examples of nonlinear transmission lines are
Josephson transmission lines (JTL) \cite{barone,pedersen,tinkham,kadin}, constructed from Josephson junctions (JJs), capacitors, and (possibly) resistors.
The unique nonlinear properties of JTL allow to construct
soliton propagators,
microwave oscillators, mixers, detectors,
parametric amplifiers, and  analog amplifiers \cite{pedersen,kadin,tinkham}.

Transmission lines formed by JJs  connected in series were
studied beginning from  1990s, though much less than transmission lines
formed by JJs  connected in parallel \cite{solitons}.
However, the former began to attract quite a lot of attention  recently \cite{yaakobi,brien,macklin,kochetov,zorin,basko,dixon,goldstein}, especially
in connection with possible JTL traveling wave parametric amplification
\cite{white,miano,pekker}.

The interest in studies of discrete nonlinear electrical transmission lines, in particular of lossy nonlinear transmission lines, has started some time ago \cite{rosenau,chen,mohebbi}, but it became even more pronounced recently
\cite{ricketts,houwe,katayama,sekulic}.
These studies should be seen in the general context of waves in strongly nonlinear
discrete systems \cite{kevrikidis0,english,kevrikidis,nesterenko0,malomed2,nesterenko,malomed}.
A very informative and very recent review  of nonlinear electric transmission
networks one can find in Ref. \cite{malomed3}.

In our previous publications we considered  travelling waves in the continuous \cite{kogan2}
and the
discrete \cite{kogan} JTL.
We have shown that such travelling waves are the kinks and the solitons in the lossless line and the shocks in the lossy line.
In the present paper we rederive part of the results obtained previously in a more concise and physically appealing way and also generalize them.
We must add that both previously and now our lossless JTL didn't (doesn't) contain the capacitor shunting the JJ,
 and our lossy JTL -- did, which made (makes) the model of the JJ more realistic.

The rest of the paper is constructed as follows.
In Section~\ref{lossless} we write down the Kirchhoff equations for the discrete  JTL, both lossless and lossy, and calculate the   velocity of shock, existing in the  latter case.
In Section~\ref{general}, on top of the continuum approximation, we  formulate the simple wave approximation,
which   decouples  two coupled  continuum JTL equations into two separate equations for the right- and left-going waves.
In  Section~\ref{again} we formulate the quasi-discrete approximation and show that the only compact travelling waves in the lossless JTL are the kinks and the solitons and calculate their velocities.
In Section \ref{yy} we  formulate the simple wave approximation to the quasi-continuum JTL equations.
We conclude in Section \ref{concl}.
In the Appendix \ref{k} we
write down the Hamiltonian, describing the lossless JTL.
In the Appendix \ref{j} we present graphical analysis of the shocks.
In the Appendix \ref{y} we use Riemann method of characteristics to justify (up to a certain degree) the simple wave approximation.

\section{The discrete JTL}
\label{lossless}

\subsection{The lossless line}
\label{loss}

\begin{figure}[h]
\includegraphics[width=1\columnwidth]{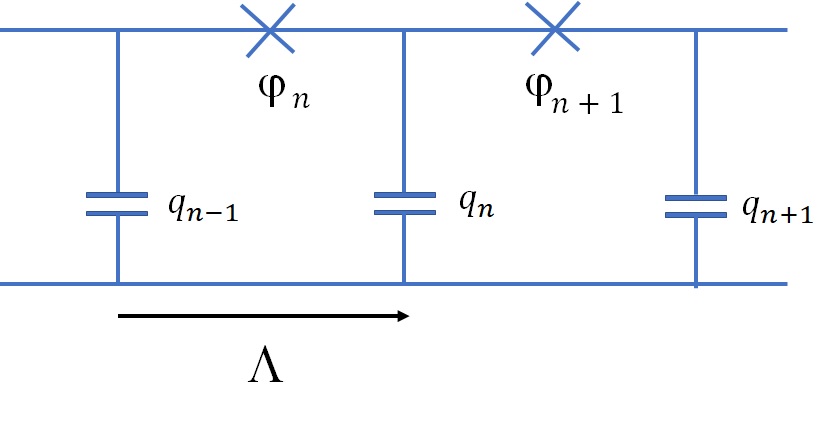}
\vskip -.5cm
\caption{Lossless discrete Josephson transmission line}
\label{trans100}
\end{figure}

Consider
the discrete model of lossless line, constructed from identical JJ and capacitors, which is shown on Fig. \ref{trans100}.
We take
as dynamical variables  the phase differences (which we for brevity will call just phases) $\varphi_n$ across the  JJs
and the charges $q_n$ of the ground capacitors. (Note that the latter choice is different from what was done in our previous publications and in the Appendix~\ref{k}.)
The  circuit equations are
\begin{subequations}
\label{ave7h}
\begin{alignat}{4}
\frac{\hbar}{2e}\frac{d \varphi_n}{d t}&=\frac{1}{C}\left(q_{n-1}-q_{n}\right) ,\label{ave7ha}\\
\frac{dq_n}{dt} &=   I_c\left(\sin\varphi_n-\sin\varphi_{n+1}\right),\label{ave7hb}
\end{alignat}
\end{subequations}
where    $C$ is the capacitance, and  $I_c$ is the critical current of the JJ.

The compact waves we are considering are characterised by the boundary conditions
\begin{eqnarray}
\label{granub}
\lim_{n\to -\infty}\varphi=\varphi_2\;,\hskip 1cm
\lim_{n\to +\infty}\varphi=\varphi_1,\\
\lim_{n\to -\infty}q=q_2\;,\hskip 1cm
\lim_{n\to +\infty}q=q_1\,.
\end{eqnarray}
(These conditions will remain the same  after we include the losses.)
Summing up  (\ref{ave7h}) from the far  left  to the far  right  we obtain
\begin{subequations}
\label{ave77}
\begin{alignat}{4}
\frac{\hbar}{2e}\sum_n\frac{d\varphi_n}{dt}&=\frac{1}{C}\left(q_2-q_1\right) ,\label{ave77a}\\
\sum_n\frac{dq_n}{dt}&=   I_c\left(\sin\varphi_2-\sin\varphi_1\right).\label{ave77b}
\end{alignat}
\end{subequations}

The travelling wave solutions satisfy equations
\begin{subequations}
\label{trat}
\begin{alignat}{4}
\varphi_n(t)&=\varphi(Ut-n\Lambda),\\
q_n(t)&=q(Ut-n\Lambda),
\end{alignat}
\end{subequations}
where $\Lambda$ is the JTL period and $U$ is the travelling wave velocity.
Hence keeping only the main harmonic in the Poisson summation formula for the sums in the l.h.s. of Eqs. (\ref{ave7h}), we can write down
\begin{subequations}
\label{e77}
\begin{alignat}{4}
\sum_n\frac{d\varphi_n}{dt}&=\frac{U}{\Lambda}\left(\varphi_2-\varphi_1\right) ,\label{e77a}\\
\sum_n\frac{dq_n}{dt}&=\frac{U}{\Lambda}\left(q_2-q_1\right),\label{e77b}
\end{alignat}
\end{subequations}
and (\ref{ave77}) becomes
\begin{subequations}
\label{ave78}
\begin{alignat}{4}
\frac{\hbar U}{2e\Lambda}\left(\varphi_2-\varphi_1\right)&=\frac{1}{C}\left(q_2-q_1\right)  ,\label{ave78a}\\
\frac{U}{\Lambda}\left(q_2-q_1\right)&= I_c\left(\sin\varphi_2-\sin\varphi_1\right),\label{ave78b}
\end{alignat}
\end{subequations}
Excluding $q_2-q_1$  we obtain
\begin{eqnarray}
\label{city}
\overline{U}^2\left(\varphi_1-\varphi_2\right)=\sin\varphi_1-\sin\varphi_2;
\end{eqnarray}
for any velocity $V$ in this paper
\begin{eqnarray}
\overline{V}\equiv\frac{\sqrt{L_J C}}{\Lambda}V.
\end{eqnarray}
where $L_J=\hbar/(2eI_c)$.
The velocity $U$ should be compared with is the  velocity $u(\varphi)$ of propagation of small amplitude disturbances on a homogeneous background $\varphi$, which is given by the equation \cite{kogan2}
\begin{eqnarray}
\overline{u}(\varphi)=\sqrt{\cos\varphi}.
\end{eqnarray}
 (Everywhere in this paper
we consider only the case $-\pi/2<\varphi<\pi/2$.)

We must admit that, strictly speaking, ignoring of the higher harmonics in the Poisson summation formula which led to (\ref{e77}) can be justified only
when the space scale of the solutions is much larger than $\Lambda$. In Ref. \cite{kogan} it was shown that this is true for weak waves (the definition of the weak wave see below).
However further on we'll forget about this limitation.

Differentiating (\ref{ave7ha}) with respect to $t$ and substituting $dq_n/dt$  from Eq. (\ref{ave7hb}), we obtain closed equation for $\varphi_n$
\begin{eqnarray}
\label{com}
\frac{d^2 \varphi_n}{d \tau^2}=\sin\varphi_{n+1}-2\sin\varphi_{n}+\sin\varphi_{n-1},
\end{eqnarray}
where we  introduced the dimensionless  time $\tau=t/\sqrt{L_J C}$.
Equation (\ref{com}) will be used in Section \ref{again}.

\subsection{The lossy line}

\begin{figure}[h]
\includegraphics[width=\columnwidth]{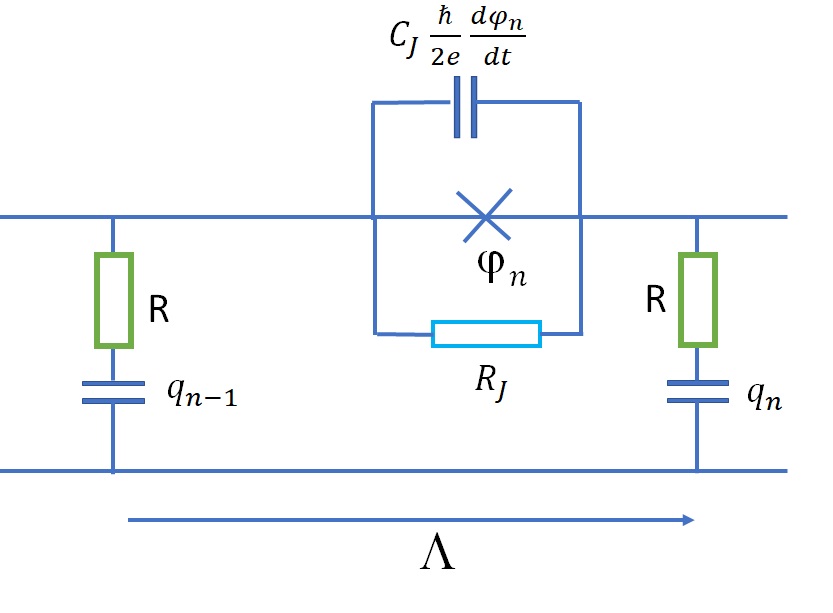}
\vskip -.5cm
\caption{Discrete Josephson transmission line with the capacitor  and the resistor shunting the JJ and another resistor in series with the ground capacitor}
 \label{trans4}
\end{figure}

Now let us consider the  JTL with the capacitors  and resistors shunting the JJs and (resistors) in series with the ground capacitors,
shown  on Fig. \ref{trans4}.
As the result,
 Eq. (\ref{ave7h}) changes to
\begin{subequations}
\label{ave8}
\begin{alignat}{4}
\frac{\hbar}{2e}\frac{d \varphi_n}{d t}&=\left(\frac{1}{C}+R\frac{d}{d t}\right)\left(q_{n-1}-q_{n}\right)  ,\label{ave8a}\\
\frac{dq_n}{dt} &=  I_c\sin\varphi_n- I_c\sin\varphi_{n+1}\nonumber\\
&+\left(\frac{\hbar}{2eR_J}\frac{d}{d t}
+C_J\frac{\hbar}{2e}\frac{d^2}{d t^2}\right)\left(\varphi_{n}-\varphi_{n+1}\right),\label{ave8b}
\end{alignat}
\end{subequations}
where $R$ is the ohmic resistor  in series with the ground
capacitor, and $C_J$ and $R_J$ are the capacitor and the ohmic resistor shunting the JJ.

Summing up  (\ref{ave8}) from the far  left to the shock   to the far  right,  we obtain instead of (\ref{ave78})
\begin{subequations}
\label{e78}
\begin{alignat}{4}
\frac{\hbar U}{2e\Lambda}\left(\varphi_1-\varphi_2\right)&=\left(\frac{1}{C}+R\frac{d}{d t}\right)\left(q_2-q_1\right),\label{e78a}\\
\frac{U}{\Lambda}\left(q_1-q_2\right)&=I_c\left(\sin\varphi_2-\sin\varphi_1\right)
\nonumber\\
+   \left(\frac{\hbar}{2eR_J}\frac{d}{d t}
\right.&+\left.C_J\frac{\hbar}{2e}\frac{d^2}{d t^2}\right)\left(\varphi_2-\varphi_1\right).
\label{e78b}
\end{alignat}
\end{subequations}
All the time derivatives are equal to zero, and we recover equations  (\ref{ave78}),
which are presented graphically in the Appendix~\ref{j}.
From (\ref{ave78}) follows (\ref{city}),
hence we obtain for the shock velocity equation \cite{kogan2,kogan}
\begin{eqnarray}
\label{velocity3}
\overline{U}^2_{\text{sh}}(\varphi_1,\varphi_2)=\frac{\sin\varphi_1-\sin\varphi_2}
{\varphi_1-\varphi_2}.
\end{eqnarray}
The situation is similar to that in fluid dynamics \cite{landau2}: the losses are necessary for the existence of the shock but don't influence the shock velocity.

The resistance in series with the ground capacitor doesn't have any clear cut physical realization, so from now on we'll put $R=0$. A more realistic model would include resistance parallel to the capacitor, but we postpone such generalization  until later time.

\section{The continuum approximation}
\label{general}

\subsection{The lossless line}

Let us write down Eq. (\ref{ave7h}) in the continuum approximation, that is  let us  treat $n$  as the continuous variable $Z$ and approximate the finite differences in the r.h.s. of the equations as the first derivative with respect to $Z$.  After which (\ref{ave7h}) takes the form
\begin{subequations}
\label{ave9}
\begin{alignat}{4}
\frac{\partial \varphi}{\partial\tau}&= -\frac{\partial Q}{\partial Z},\label{ave9a}\\
\frac{\partial Q}{\partial\tau} &=  -\cos\varphi\frac{\partial \varphi}{\partial Z},\label{ave9b}
\end{alignat}
\end{subequations}
where we  introduced  the dimensionless charge $Q=q/(I_c\sqrt{L_J C})$.

\subsubsection{The simple wave approximation}

The simple wave approximation  is decoupling of the wave equation into two separate equations for the right- and left-going waves.
To achieve that aim
let us assume that  $Q$ is a function of $\varphi$. Then we have
\begin{subequations}
\label{e2}
\begin{alignat}{4}
\frac{\partial Q}{\partial Z}&= \frac{d Q}{d \varphi}\frac{\partial \varphi}{\partial Z},\label{e2a}\\
\frac{\partial Q}{\partial\tau}&= \frac{d Q}{d \varphi}\frac{\partial \varphi}{\partial\tau}.\label{e2b}
\end{alignat}
\end{subequations}
Substituting (\ref{e2}) into (\ref{ave9}) and excluding $dQ/d\varphi$ we obtain \cite{kogan2}
\begin{eqnarray}
\label{newnew3}
\frac{\partial \varphi}{\partial \tau}
=\pm\sqrt{\cos\varphi}\frac{\partial \varphi}{\partial Z}.
\end{eqnarray}
On the other hand, excluding the partial derivatives after the substitution we obtain
\begin{eqnarray}
\label{qq}
\left(\frac{d Q}{d \varphi}\right)^2=\cos\varphi\Rightarrow
Q=\pm\int \sqrt{\cos\varphi}d\varphi.
\end{eqnarray}
Another line of reasoning which can lead from to (\ref{ave9}) to (\ref{newnew3}) and (\ref{qq}) is based on  Riemann method of characteristics \cite{landau2} and is presented in the Appendix \ref{y}.

\subsection{The lossy line}
\label{lo}

 Applying the
continuum approximation to (\ref{ave8})  we obtain
\begin{subequations}
\label{ve9b}
\begin{alignat}{4}
\frac{\partial \varphi}{\partial\tau}&= -\frac{\partial Q}{\partial Z},\label{ve9ba}\\
\frac{\partial Q}{\partial\tau} &=  -\frac{\partial \sin\varphi}{\partial Z}-\frac{Z_J}{R_J}\frac{\partial^2\varphi}{\partial\tau\partial Z }
-\frac{C_J}{C}\frac{\partial^3\varphi}{\partial\tau^2\partial Z},\label{ve9bb}
\end{alignat}
\end{subequations}
where $Z_J\equiv\sqrt{L_J/C}$ is the characteristic impedance of the JTL.
We can exclude $Q$ from (\ref{ve9b}) and obtain closed equation for $\varphi$
\begin{eqnarray}
\label{co}
\frac{\partial ^2\varphi}{\partial\tau^2}=\frac{\partial ^2}{\partial Z^2}
\left(\sin\varphi
+\frac{Z_J}{R_J}\frac{\partial\varphi}{\partial\tau}
+\frac{C_J}{C}\frac{\partial^2\varphi}{\partial\tau^2}\right).
\end{eqnarray}

If we expand the sine function in the r.h.s. of (\ref{co}) in Taylor series, keep the first two terms of the expansion, substitute $\varphi=\partial \widetilde{\varphi}/\partial Z$ and integrate the resulting equation with respect to $Z$ we obtain
\begin{eqnarray}
\label{coco}
\frac{\partial ^2\widetilde{\varphi}}{\partial\tau^2}
=\frac{\partial^2\widetilde{\varphi}}{\partial Z^2}
-\frac{1}{6}\frac{\partial}{\partial Z}\left(\frac{\partial\widetilde{\varphi}}{\partial Z}\right)^3
+\frac{Z_J}{R_J}\frac{\partial^3\widetilde{\varphi}}{\partial\tau\partial Z^2}
+\frac{C_J}{C}\frac{\partial^4\widetilde{\varphi}}{\partial\tau^2\partial Z^2},\nonumber\\
\end{eqnarray}
which for $R_J=\infty$ exactly coincides with Eq. (22) from the seminal paper \cite{yaakobi} for $R=0$.
Note that (\ref{coco}), apart from being an approximation to (\ref{co}), corresponds to a different choice of the dynamical variables. While in our approach, $\varphi_n$ is the phase difference across the appropriate JJ, in the approach of Ref. \cite{yaakobi}, $\widetilde{\varphi}_n$ corresponds to the Josephson phase  in between the JJs (Eq. (12) of the reference).

\subsubsection{The simple wave approximation}

Substituting (\ref{e2}) into (\ref{ve9b}) and excluding $dQ/d\varphi$ we obtain
\begin{eqnarray}
\label{gh}
\left(\frac{\partial \varphi}{\partial\tau}\right)^2=\cos\varphi\left(\frac{\partial \varphi}{\partial Z}\right)^2
+\frac{\partial \varphi}{\partial Z}\left(\frac{Z_J}{R_J}
\frac{\partial^2\varphi}{\partial\tau\partial Z }+\frac{C_J}{C}
\frac{\partial^3\varphi}{\partial\tau^2\partial Z}\right).\nonumber\\
\end{eqnarray}
If the higher derivative terms in (\ref{gh}) is only a small correction,  we can approximately extract square root from both sides of the equation and obtain
\begin{eqnarray}
\label{ve9}
\frac{\partial \varphi}{\partial\tau}=\pm\left(\sqrt{\cos\varphi}\frac{\partial \varphi}{\partial Z}+\frac{Z_J}{2R_J\sqrt{\cos\varphi}}
\frac{\partial^2\varphi}{\partial\tau\partial Z }\right.\nonumber\\
\left.
+\frac{C_J}{2C\sqrt{\cos\varphi}}
\frac{\partial^3\varphi}{\partial\tau^2\partial Z}\right).
\end{eqnarray}
In the framework of out  cavalier treatment, we may as well  modify  Eq. (\ref{ve9}) to
\begin{eqnarray}
\label{e9b}
\frac{\partial \varphi}{\partial\tau}=\pm\sqrt{\cos\varphi}\left(\frac{\partial \varphi}{\partial Z}+\frac{C_J}{2C}
\frac{\partial^3\varphi}{\partial  Z^3}\right)+\frac{Z_J}{2R_J}
\frac{\partial^2\varphi}{\partial Z^2 }.
\end{eqnarray}
We can keep Eq.(\ref{qq}) for both versions.

\subsubsection{The travelling waves}
\label{trtr}

Consider
the travelling waves, which satisfy equation
\begin{eqnarray}
\label{trat2}
\varphi(\tau,Z)=\varphi(x),
\end{eqnarray}
where  $x=\overline{U}\tau-Z$.
Making the  ansatz in (\ref{ve9b}), we obtain  (after an integration) \cite{kogan2}
\begin{eqnarray}
\label{velo}
\frac{C_J\overline{U}^2}{C}\frac{d^2\varphi}{d x^2 }+\frac{Z_J\overline{U}}{R_J}\frac{d\varphi}{d x }
=\overline{U}^2\varphi-\sin\varphi+F,
\end{eqnarray}
where $F$ is the constant of integration.
On the other hand,
from Eq. (\ref{ve9}) we obtain the equation
\begin{eqnarray}
\label{veloc2}
\frac{C_J\overline{U}^2}{2C}\frac{d^3\varphi}{dx^3}
+\frac{Z_J\overline{U}}{2R_J}\frac{d^2\varphi}{d x^2 }
=\left[|\overline{U}|\sqrt{\cos\varphi}-\cos\varphi\right]
\frac{d \varphi}{dx},
\end{eqnarray}
and then, after the integration, the equation
\begin{eqnarray}
\label{eloc2}
\frac{C_J\overline{U}^2}{2C}\frac{d^2\varphi}{dx^2}
+\frac{Z_J\overline{U}}{2R_J}\frac{d\varphi}{d x}
=|\overline{U}|\int\sqrt{\cos\varphi}d \varphi-\sin\varphi.
\end{eqnarray}
Both Eq. (\ref{velo}) and Eq. (\ref{eloc2}) describe damped motion of the fictitious Newtonian particle in the potential well, with $x$ playing the role of time.

\subsubsection{The shocks}
\label{pata}

Let us consider for the sake of definiteness the case $\overline{U}>0$.
The shocks correspond to the boundary conditions
\begin{eqnarray}
\varphi(-\infty)=\varphi_1,\hskip 1cm \varphi(+\infty)=\varphi_2.
\end{eqnarray}
The point $\varphi_1$ corresponds to the unstable equilibrium,  and the point $\varphi_2$ -- to the stable.

Let us start from  Eq. (\ref{velo}). The r.h.s. of the equation for $\varphi=\varphi_1$ is equal to that for $\varphi=\varphi_2$
(they are both equal to zero).
Hence we  recover (\ref{velocity3}).
Differentiating the r.h.s. of (\ref{velo}) with respect to $\varphi$ we obtain
\begin{subequations}
\label{sho}
\begin{alignat}{4}
\overline{U}^2-\cos\varphi_1&>0,\label{shoa}\\
\overline{U}^2-\cos\varphi_2&<0.\label{shob}
\end{alignat}
\end{subequations}
The inequalities (\ref{sho}) reflect the  well known  in the nonlinear waves theory  fact: the shock velocity is smaller than the sound velocity   in the region behind the shock
but larger than the sound velocity in the region before the shock  \cite{whitham}.

To get some confidence in the simple wave approximation,
we should check up whether and to what extent (\ref{velocity3}) and (\ref{sho})  follow from  the  approximate equation (\ref{eloc2}).
Differentiating the r.h.s. of (\ref{eloc2}) with respect to $\varphi$ we obtain
\begin{subequations}
\label{ho}
\begin{alignat}{4}
|\overline{U}|\sqrt{\cos\varphi_1}-\cos\varphi_1&>0,\label{hoa}\\
|\overline{U}|\sqrt{\cos\varphi_2}-\cos\varphi_2&<0,\label{hob}
\end{alignat}
\end{subequations}
which is equivalent to (\ref{sho})).
Writing down  (\ref{eloc2}) for $\varphi_1$ and $\varphi_2$ and taking into account  the boundary conditions, we obtain
\begin{eqnarray}
\label{velocb}
\overline{U}_{\text{sh}}(\varphi_1,\varphi_2)=
\frac{\sin\varphi_1-\sin\varphi_2}{\int^{\varphi_1}_{\varphi_2}\sqrt{\cos\varphi} d\varphi}.
\end{eqnarray}

If we use the travelling wave ansatz for (\ref{ve9b}) and
consider the particular case $C_J=0$, the resulting equation can also be easily integrated and we obtain
\begin{eqnarray}
\label{veloc}
\overline{U}_{\text{sh}}(\varphi_1,\varphi_2)=
\frac{\int^{\varphi_1}_{\varphi_2}\sqrt{\cos\varphi} d\varphi}{\varphi_1-\varphi_2}.
\end{eqnarray}
Equations (\ref{ve9}) and (\ref{e9b})  differ only in the higher derivative terms. Looking at Eqs. (\ref{velocb}) and (\ref{veloc})  we realize that
in both cases the shock velocity doesn't depend upon the coefficients before the  derivatives.
However,  different structure of the terms  leads  to different shock velocities.

\begin{figure}[h]
\includegraphics[width=.7\columnwidth]{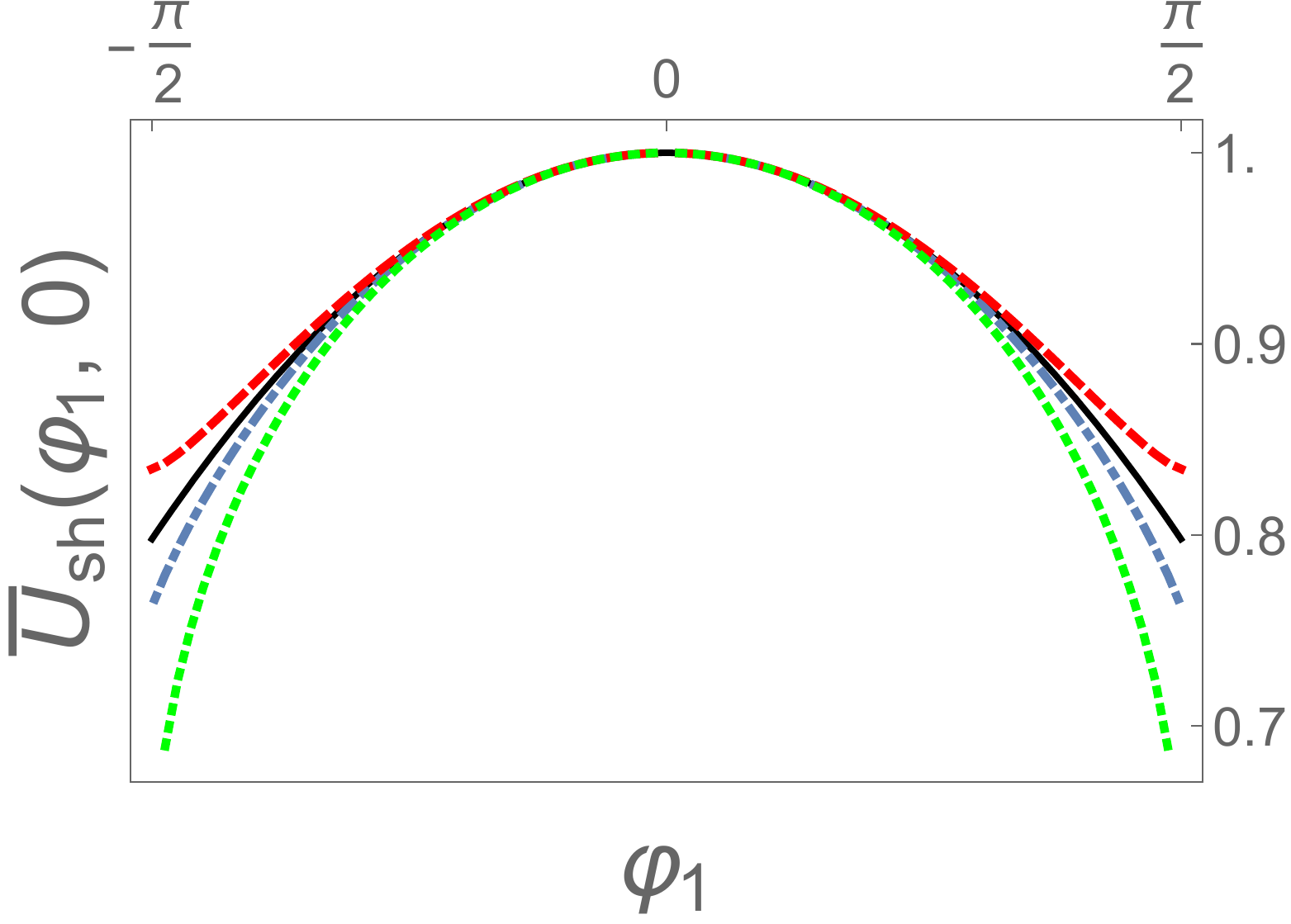}
\caption{The shock velocity $\overline{U}_{\text{sh}}(\varphi_1,\varphi_2)$ for $\varphi_2=0$ as function of $\varphi_1$, as given by Eq. (\ref{velocity3}) (solid line),
Eq. (\ref{velocb}) (red dashed line) and Eq. (\ref{veloc})  (blue dot-dashed line). The green dotted line corresponds to Eq. (\ref{hru2}) (see Section \ref{patash}).}
\label{compar}
\end{figure}

We  plotted the shock velocity $\overline{U}_{\text{sh}}(\varphi_1,\varphi_2)$, as given by Eqs. (\ref{velocity3}),  (\ref{velocb}) and (\ref{veloc}),
as function of one  phase, the other being fixed,   in Fig. \ref{compar}.
It is nice to realize, that though looking differently, all the equations give close results.

\section{The quasi-discrete approximation}
\label{again}

We stopped our consideration of the lossless line at
Eq. (\ref{com}). Now let us return to that equation.
Considering $\varphi$ as functions of the continuous variable $Z$ (instead of the discrete variable $n$) and making Taylor expansion of the r.h.s. of (\ref{com})  we obtain \cite{kogan}
\begin{eqnarray}
\label{old}
\frac{\partial^2 \varphi}{\partial \tau^2}
=\sum_{m=1}^{\infty}\frac{2}{(2m)!}\frac{\partial^{2m}\sin\varphi}{\partial Z^{2m}}.
\end{eqnarray}

\subsection{The travelling waves:  the kinks and the solitons}
\label{ks}

Using the travelling wave ansatz  we arrive from (\ref{old}) to
\begin{eqnarray}
\label{v7}
\overline{U}^2\frac{d^2\varphi}{d x^2}
=\sum_{m=1}^{\infty}\frac{2}{(2m)!}\frac{d^{2m}\sin\varphi}{d x^{2m}}.
\end{eqnarray}
Integrating  we obtain
\begin{eqnarray}
\label{v772}
\overline{U}^2\frac{d\varphi}{d x}
=\sum_{m=1}^{\infty}\frac{2}{(2m)!}\frac{d^{2m-1}\sin\varphi}{d x^{2m-1}}
\end{eqnarray}
(the  constant of integration is equal to zero because of the boundary conditions).
Integrating the second time we get
\begin{eqnarray}
\label{v77}
\sum_{m=1}^{\infty}\frac{2}{(2m+2)!}\frac{d^{2m}\sin\varphi}{d x^{2m}}&=&\overline{U}^2\varphi-\sin\varphi +F\nonumber\\
&\equiv& -\frac{d\Pi(\sin\varphi)}{d\sin\varphi},
\end{eqnarray}
where $F$ is the constant of integration and
\begin{eqnarray}
\label{v10b}
\Pi(\sin\varphi)=-\overline{U}^2(\varphi\sin\varphi+\cos\varphi)
+\frac{1}{2}\sin^2\varphi-F\sin\varphi.\nonumber\\
\end{eqnarray}
Taking into account the boundary conditions (\ref{granub}), we obtain from (\ref{v77})
\begin{eqnarray}
\label{v797}
\overline{U}^2\varphi_{1,2}-\sin\varphi_{1,2} +F=0.
\end{eqnarray}
Equation (\ref{city}) obviously follows from (\ref{v797}).

However for the lossless JTL, in distinction from the lossy case, there exists additional conservation law,
which can be obtained if we multiply  (\ref{v772}) by  $d\sin\varphi/dx$ and   integrate once again from $-\infty$ to $+\infty$.  From the  identities
\begin{eqnarray}
\label{r}
\frac{d^{2m}y}{d x^{2m}}\frac{dy}{d x}=\frac{d}{dx}\left[\frac{d^{2m-1}y}{d x^{2m-1}}\frac{dy}{d X}-\frac{d^{2m-2}y}{d x^{2m-2}}\frac{d^2y}{d x^2}
\right.\nonumber\\
+\left.\frac{d^{2m-3}y}{d x^{2m-3}}\frac{d^3y}{d x^3}-\dots-\frac{1}{2}(-1)^m\left(\frac{d^my}{d x^m}\right)^2\right]
\end{eqnarray}
follows that the integral from the l.h.s. is equal to zero, and we obtain
\begin{eqnarray}
\label{v10}
\Pi(\sin\varphi_1)=\Pi(\sin\varphi_2).
\end{eqnarray}
Comparing (\ref{v797}) and (\ref{v10}) we realize that there are two possibilities for the compact travelling waves we are considering: $\varphi_2=-\varphi_1$ and $\varphi_2=\varphi_1$. In the former case we obtain kink, in the latter -- soliton  \cite{kogan}.
For the kink $F=0$ and
\begin{eqnarray}
\label{velocitk}
\overline{U}^2_{\text{kink}}(\varphi_1)=\frac{\sin\varphi_1}{\varphi_1}.
\end{eqnarray}

\section{The  quasi-continuum approximation}
\label{yy}

In this Section we continue to study the properties of the lossless JTL.
In the continuum approximation  we kept in the sum in the r.h.s. of (\ref{old}) only the first term.
In the quasi-continuum approximation  we keep   the two first terms, thus obtaining
\begin{eqnarray}
\label{oldnew2}
\frac{\partial^2 \varphi}{\partial \tau^2}
=\frac{\partial^2\sin\varphi}{\partial Z^2}+\frac{1}{12}\frac{\partial^4\sin\varphi}{\partial Z^4}.
\end{eqnarray}
Just a side remark.
Equation  (\ref{co}) in the quasi-continuum approximation  becomes
\begin{eqnarray}
\frac{\partial ^2\varphi}{\partial\tau^2}=\left(\frac{\partial ^2}{\partial Z^2}+\frac{1}{12}\frac{\partial ^4}{\partial Z^4}\right)
\left(\sin\varphi
+\frac{Z_J}{R_J}\frac{\partial\varphi}{\partial\tau}
+\frac{C_J}{C}\frac{\partial^2\varphi}{\partial\tau^2}\right).\nonumber\\
\end{eqnarray}

\subsection{The kinks and the solitons}

In the quasi-continuum approximation  (\ref{v77})  is truncated to
 \begin{eqnarray}
\label{v774}
\frac{1}{12}\frac{d^{2}\sin\varphi}{d x^{2}}=-\frac{d\Pi(\sin\varphi)}{d\sin\varphi}.
\end{eqnarray}
Thus after the truncation we obtained Newtonian equation describing motion of the fictitious particle in the potential well $\Pi(\sin\varphi)$, with $\sin\varphi$ playing the role of the coordinate and $x$ playing the role of time
 \cite{kogan}. Only now, in distinction from what is described in Section \ref{pata}, the motion is undamped. Hence it starts in the infinite past at the maximum of the potential and ends in the infinite future again at the maximum. It can be a different maximum, or it can be the same one.
The condition $\varphi_1=\pm\varphi_2$ receives a very simple interpretation: because of the particle energy conservation, maximum of the well potential at $\varphi=\varphi_1$ should be equal to that at $\varphi=\varphi_2$.

\subsubsection{The soliton velocity}

For the soliton, two equations (\ref{v797}) turn into the single equation, and we need an additional one.
To get this equation, let us  introduce the
additional parameter -- the amplitude of the soliton  (maximally different from $\varphi_1$ value of $\varphi$), which we will  designate as $\varphi_0$.
From the energy conservation law we obtain
\begin{eqnarray}
\label{v106}
\Pi(\sin\varphi_1)=\Pi(\sin\varphi_0).
\end{eqnarray}
From Eqs. (\ref{v797}) and (\ref{v106}) we can obtain the velocity of the soliton \cite{kogan}:
\begin{eqnarray}
\overline{U}^2_{\text{sol}}(\varphi_1,\varphi_0)
=\frac{\left(\sin\varphi_1-\sin\varphi_0\right)^2}
{2[\cos\varphi_0-\cos\varphi_1-(\varphi_1-\varphi_0)\sin\varphi_0]}.\nonumber\\
\end{eqnarray}

\subsection{Weak waves}
\label{w}

Equation (\ref{oldnew2}) can be simplified for weak waves, characterized by the inequality $|\varphi-\varphi_1|\ll 1$, where $\varphi_1$ is some constant quantity.
In this case it is convenient to introduce the new variable $\psi$ by the equation
$\varphi=\varphi_1+\psi$. Expanding Eq. (\ref{old})  with respect to $\psi$, keeping only the two lowest order terms of the expansion  and ignoring in addition the nonlinear term which contains the forth order derivative, we obtain:   for $\varphi_1\sim 1$ -- the Boussinesq equation
\begin{eqnarray}
\label{old8}
\frac{\partial^2 \psi}{\partial \tau^2}
=\cos\varphi_1\frac{\partial^2\psi}{\partial Z^2}-\frac{\sin\varphi_1}{2}\frac{\partial^2\psi^2}{\partial Z^2}
+\frac{\cos\varphi_1}{12}\frac{\partial^4\psi}{\partial Z^4},
\end{eqnarray}
and for $\varphi_1=0$ -- the modified Boussinesq equation
\begin{eqnarray}
\label{old55}
\frac{\partial^2 \psi}{\partial \tau^2}
=\frac{\partial^2\psi}{\partial Z^2}-\frac{1}{6}\frac{\partial^2\psi^3}{\partial Z^2}
+\frac{1}{12}\frac{\partial^4\psi}{\partial Z^4}.
\end{eqnarray}

Note that if  we are considering modulated high frequency harmonic wave \cite{kogan3}, the quadratic with respect to $\psi$ term  in the expansion of Eq. (\ref{oldnew2}) is irrelevant (in the lowest order of perturbation theory) and can be omitted, independently of the value of $\varphi_1$, \cite{solitons} but the cubic term should be kept.

\subsection{The simple wave approximation}
\label{patash}

Let us return to Eq. (\ref{ave7h}). Let us  treat $n$ in the latter as the continuous variable $Z$, expand the finite difference in the r.h.s. of the equations in Taylor series around $n\pm 1/2$  and keep in each case  the first two terms of the  expansion, after which the equation takes the form
\begin{subequations}
\label{old5}
\begin{alignat}{4}
\frac{\partial \varphi}{\partial\tau}&= -\left(\frac{\partial }{\partial Z}
+\frac{1}{24}\frac{\partial^3 }{\partial Z^3}\right)Q,\label{old5a}\\
\frac{\partial Q}{\partial\tau}& =  -\left(\frac{\partial }{\partial Z}
+\frac{1}{24}\frac{\partial^3 }{\partial Z^3}\right)\sin\varphi.\label{old5b}
\end{alignat}
\end{subequations}
Note that excluding $Q$ from (\ref{old5}) and ignoring in the resulting equation the sixth order derivative we recover (\ref{oldnew2}).

Let us transform (\ref{old5})  assuming that  $Q$ is a function of $\varphi$
and, in addition,  approximating the third order partial derivative terms as it is presented below
\begin{subequations}
\label{old6}
\begin{alignat}{4}
\frac{\partial \varphi}{\partial\tau}&= -\frac{d Q}{d \varphi}\left(\frac{\partial }{\partial Z}
+\frac{1}{24}\frac{\partial^3 }{\partial Z^3}\right)\varphi,\label{old6a}\\
\frac{d Q}{d \varphi}\frac{\partial \varphi}{\partial\tau} &=  -\cos\varphi\left(\frac{\partial }{\partial Z}
+\frac{1}{24}\frac{\partial^3 }{\partial Z^3}\right)\varphi.\label{old6b}
\end{alignat}
\end{subequations}
Excluding $dQ/d\varphi$ we obtain
\begin{eqnarray}
\label{newnew222}
\frac{\partial \varphi}{\partial \tau}
=\pm \sqrt{\cos\varphi}\left(\frac{\partial\varphi}{\partial Z}
+\frac{1}{24}\frac{\partial^3\varphi}{\partial Z^3}\right).
\end{eqnarray}

When the space scale  of a wave is much larger than $\Lambda$ we can ignore the third order derivative in (\ref{newnew222}), which brings us back to Eq. (\ref{newnew3}).
However, when studying kinks and solitons, with their space scale of order of $\Lambda$, the third order derivative is of crucial importance.

For the travelling waves from (\ref{newnew222}) we obtain
\begin{eqnarray}
\label{ew}
\frac{1}{24}\frac{d^3\varphi}{dx^3}=\left[\frac{|\overline{U}|}{\sqrt{\cos\varphi}}-1\right]
\frac{d\varphi}{d x}
\end{eqnarray}
and, after the integration,
\begin{eqnarray}
\label{e7}
\frac{1}{24}\frac{d^2\varphi}{dx^2}=|\overline{U}|\int\frac{d\varphi}{\sqrt{\cos\varphi}}
-\varphi+F
\end{eqnarray}
where $F$ is the constant of integration (compare with (\ref{v774})).
We again obtained Newtonian equation describing  motion of the fictitious Newtonian particle in the potential well.

After we  impose  the  boundary conditions (\ref{granub}),
the r.h.s. of (\ref{e7}) for $\varphi=\varphi_1$ becomes equal to that for $\varphi=\varphi_2$
(they are both equal to zero).
So we obtain
(keeping in mind the existence of the losses) the result for the shock velocity
\begin{eqnarray}
\label{hru2}
\overline{U}_{\text{sh}}(\varphi_1,\varphi_2)
=\frac{\varphi_1-\varphi_2}
{\int_{\varphi_2}^{\varphi_1}d\varphi/\sqrt{\cos\varphi}},
\end{eqnarray}
which turns out to be numerically quite close to (\ref{city}) (see Fig. \ref{compar}).

On the other hand, in the absence of losses there is  additional conservation law. If we present the r.h.s. of (\ref{e7}) as
\begin{eqnarray}
|\overline{U}|\int\frac{d\varphi}{\sqrt{\cos\varphi}}
-\varphi\equiv-\frac{d\Pi^{swa}(\varphi)}{d\varphi},
\end{eqnarray}
we immediately understand that the fictitious particle energy conservation means
\begin{eqnarray}
\label{pp}
\Pi^{swa}(\varphi_1)=\Pi^{swa}(\varphi_2).
\end{eqnarray}
Equation (\ref{pp}), together with equality of the derivative of the potential to zero both at $\varphi=\varphi_1$ and at $\varphi=\varphi_2$,
leads to the result of Section \ref{ks}
 $\varphi_1=\pm\varphi_2$.

\subsubsection{Weak waves}

In the same approximations as in Section \ref{w}, Eq. (\ref{newnew222})
for  $\varphi_1\sim 1$ is simplified to
\begin{eqnarray}
\label{cocob}
\frac{\partial \psi}{\partial \tau}
=\pm\sqrt{\cos\varphi_1}\left(\frac{\partial\psi}{\partial Z}
-\frac{\tan\varphi_1}{2}\psi\frac{\partial\psi}{\partial Z}
+\frac{1}{24}\frac{\partial^3\psi}{\partial Z^3}\right),
\end{eqnarray}
which is the Korteweg-de Vries (KdV) equation.
On the other hand, for $\varphi_1=0$, (\ref{newnew222}) is simplified to
\begin{eqnarray}
\label{omb}
\frac{\partial \psi}{\partial \tau}
=\pm\left(\frac{\partial\psi}{\partial Z}-\frac{1}{4}\psi^2\frac{\partial\psi}{\partial Z}
+\frac{1}{24}\frac{\partial^3\psi}{\partial Z^3}\right).
\end{eqnarray}
which is the modified Korteweg-de Vries (mKdV) equation \cite{katayama}.
Note that Eqs. (\ref{cocob}) and (\ref{omb}) could have been derived  from Eqs. (\ref{old8}) and (\ref{old55}) respectively (see e.g. Ref. \cite{kamchatnov}).
This gives us additional confidence in  the simple wave approximation (\ref{newnew222}).

\subsubsection{The discreet equation for the phase}
\label{dii}

Inverting the process of discretizion which led Eq.(\ref{newnew222}),
we modify the former to
\begin{eqnarray}
\label{diz}
\frac{\partial \varphi_n}{\partial \tau}
&=\sqrt{\cos\varphi_n}\left(\varphi_{n\pm 1}-\varphi_n\right).
\end{eqnarray}
Analysis of Eq. (\ref{diz}) we postpone until later time. Right now we want to show that the equation leads to the same shock velocity as (\ref{newnew222}).

Dividing both sides of  (\ref{diz}) by $\overline{u}(\varphi_n)$ and summing up from the far  left  to the far  right  we obtain
\begin{eqnarray}
\label{hru}
\sum_n\frac{1}{\sqrt{\cos\varphi_n}}\frac{d\varphi_n}{d\tau}
=\pm\left(\varphi_2-\varphi_1\right).
\end{eqnarray}
Introducing the travelling wave ansatz and keeping only the main harmonic in the Poisson summation formula for the sum in the l.h.s. of (\ref{hru}), as it was done in Section \ref{loss}, we reproduce (\ref{hru2}).

\subsubsection{The continuum approximation regained}

Equations
(\ref{e9b}) and (\ref{newnew222})  can be combined together into a single one
\begin{eqnarray}
\label{fi}
\frac{\partial \varphi}{\partial\tau}&=&\pm\sqrt{\cos\varphi}\left[\frac{\partial \varphi}{\partial Z}+\left(\frac{C_J}{2C}
+\frac{1}{24}\right)\frac{\partial^3\varphi}{\partial Z^3}\right]\nonumber\\
&+&\frac{Z_J}{2R_J}
\frac{\partial^2\varphi}{\partial Z^2}.
\end{eqnarray}
The transition from (\ref{fi}) to (\ref{newnew222}) when $C_J/C\ll 1$ and
$Z_J/R_J\gg 1$ is obvious.  To better understand the transition to (\ref{e9b}) let us restore dimensions in Eq. (\ref{fi}), that is let us return from $\tau$ to $t$ and introduce $z=\Lambda Z$ instead of $Z$. Then we obtain
\begin{eqnarray}
\label{fid}
\frac{\sqrt{L_J C}}{\Lambda}\frac{\partial \varphi}{\partial t}&=&\pm\sqrt{\cos\varphi}\left[\frac{\partial \varphi}{\partial z}+\left(\frac{C_J\Lambda^2}{2C}
+\frac{\Lambda^2}{24}\right)\frac{\partial^3\varphi}{\partial z^3}\right]\nonumber\\
&+&\frac{\sqrt{L_J/C}\Lambda}{2R_J}
\frac{\partial^2\varphi}{\partial z^2}.
\end{eqnarray}
Now let us formally consider the limit $\Lambda\to 0$, keeping the properties of the JJ and the space scale of the wave constant,  and look at the r.h.s. of (\ref{fid}). It is obvious that  the term $\Lambda^2\partial^3\varphi/\partial Z^3$ decreases with $\Lambda$ as $\Lambda^2$.
A glance at Fig. \ref{trans4} would convince everyone that $C\sim \Lambda$, hence
the term with the second derivative would decrease with $\Lambda$ as $\sqrt{\Lambda}$ and the term
$(C_J\Lambda^2/2C)\partial^3\varphi/\partial z^3$ -- as $\Lambda$. Hence in this limit we can ignore the discrete nature of the JTL and use the continuum approximation.

\section{Conclusions}
\label{concl}

In this paper we follow the ideas of our previous publications.
We present  a more general and physically appealing, than it was before, derivation of the formulas for the compact travelling waves velocity.
We analyse in details both the continuum and the quasi-continuum approximations to the discrete JTL. On top of each of these approximations we develop  the simple wave
approximation, which
decouples  the JTL equations   into two separate equations for the right- and left-going waves.   Equations
(\ref{e9b}) and (\ref{newnew222}) and their combination (\ref{fi}), obtained in the framework of the simple wave approximation, are of the well known type \cite{whitham}.  Thus the approximation makes, in particular, studying of the shock waves formation
much easier.

\begin{acknowledgments}

We are grateful to  M. Goldstein and B. Malomed for the discussion.

\end{acknowledgments}

\begin{appendix}

\section{The Hamiltonian  of the lossless JTL}
\label{k}

\begin{figure}[h]
\includegraphics[width=\columnwidth]{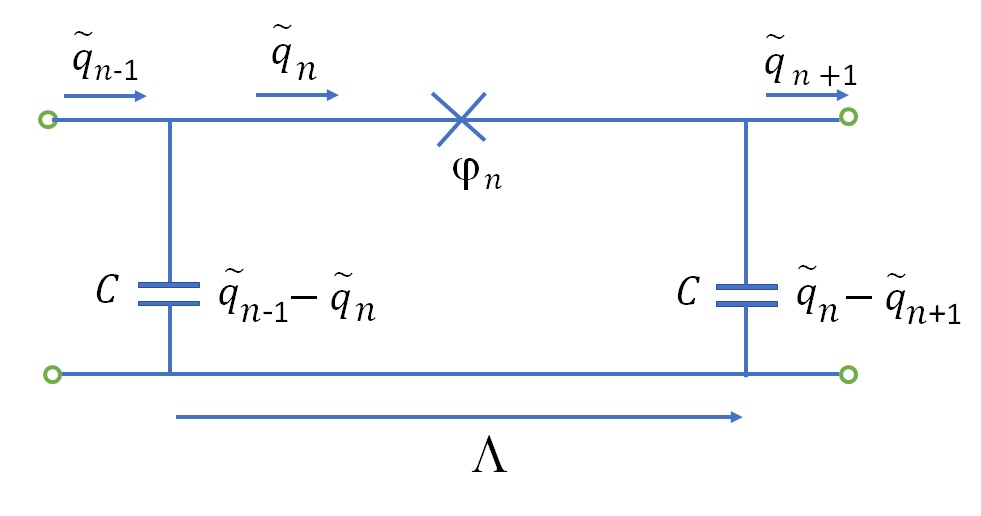}
\vskip -.5cm
\caption{Lossless discrete Josephson transmission line: different choice of the dynamical variables.}
\label{trans1000}
\end{figure}

Let us write down the JTL equations for the lossless JTL choosing as  the dynamical variables
 the phases $\varphi_n$ and the charges $\widetilde{q}_n$ which have passed through the  JJs, like it is shown on Fig. \ref{trans1000}.
The result is
\begin{subequations}
\label{ave7}
\begin{alignat}{4}
\frac{\hbar}{2e}\frac{d \varphi_n}{d t}&=\frac{1}{C}\left(\widetilde{q}_{n+1}-2\widetilde{q}_{n}+\widetilde{q}_{n-1}\right) ,\label{ave7a}\\
\frac{d\widetilde{q}_n}{dt} &=   I_c\sin\varphi_n.\label{ave7b}
\end{alignat}
\end{subequations}
The advantage of such choice of variables is that
 Eqs. (\ref{ave7}) are the canonical equations for the Hamiltonian \cite{kogan2}
\begin{eqnarray}
\label{hamm}
{\cal H}=-\frac{\hbar}{2e}I_c\sum_n\cos\varphi_n
+\frac{1}{2C}\sum_n \left(\widetilde{q}_{n}-\widetilde{q}_{n+1}\right)^2,
\end{eqnarray}
where the Poisson bracket is
\begin{eqnarray}
\left\{\widetilde{q}_{n},\varphi_{n'}\right\}\equiv\frac{2e}{\hbar}\delta_{nn'}.
\end{eqnarray}
It is interesting to compare the Hamiltonian (\ref{hamm}) with that of Fermi-Pasta-Ulam-Tsingou (FPUT) problem \cite{gallavotti}. The nonlinearity of the problem of the JTL is not due to non-quadratic potential energy, but due to the non-quadratic "kinetic" energy.

In the quasi-continuum approximation the Hamiltonian takes the form
\begin{eqnarray}
\label{hamman4}
{\cal H}=\int \left[-\frac{\hbar}{2e}I_c\cos\varphi
+\frac{1}{2C} \left(\frac{\partial \widetilde{q}}{\partial Z}\right)^2
+\frac{1}{24C} \left(\frac{\partial^2 \widetilde{q}}{\partial Z^2}\right)^2\right]dZ.\nonumber\\
\end{eqnarray}
where the Poisson bracket is
\begin{eqnarray}
\left\{\widetilde{q}(Z),\varphi(Z')\right\}\equiv\frac{2e}{\hbar}\delta(Z-Z').
\end{eqnarray}
In the continuum approximation the last term in the brackets in the integrand is absent.

\section{The Josephson curve}
\label{j}

Equation (\ref{velocity3}) is very concise and appealing, but actually equations (\ref{ave78}) are more meaningful. In fact,
in a typical situation we know not the
Josephson phases on both sides of the shock, but the phase before the shock and
the voltage difference applied (that is $Q_2-Q_1$).

\begin{figure}[h]
\includegraphics[width=.7\columnwidth]{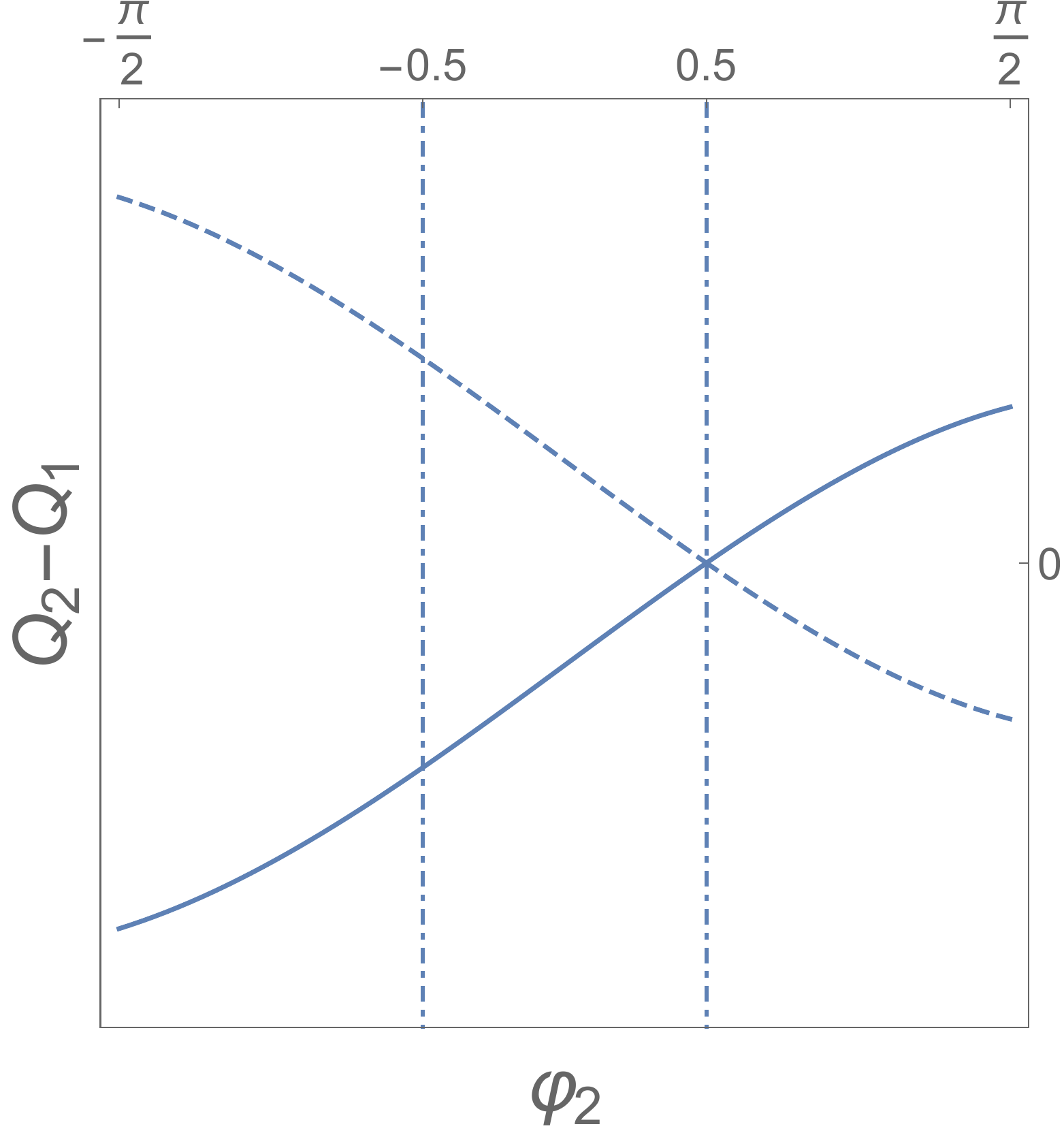}
\caption{Josephson curve given by Eq. (\ref{kompa})  for $\varphi_1=.5$. The solid curve represents the shocks moving to the right, the dashed -- to the left.}
\label{jo}
\end{figure}

To find the phase after the shock,
let us divide (\ref{ave78a}) to (\ref{ave78b}) to obtain
\begin{eqnarray}
\label{kompa}
\left(\varphi_2-\varphi_1\right)\left(\sin\varphi_2-\sin\varphi_1\right)
=\left(Q_2-Q_1\right)^2.
\end{eqnarray}
Being plotted in the coordinates $\varphi_1$, $\varphi_2$ and $Q_2-Q_1$, Eq. (\ref{kompa}) defines a surface.
We'll call
the curve  given by (\ref{kompa}) for one phase, say $\varphi_1$, fixed -- the Josephson curve.
Such curve for $\varphi_1=.5$ is presented in Fig. \ref{jo}.
From inspection of (\ref{ave78}) one realises that the solid branch  describes the shocks moving to the right, $\varphi_1$ being the phase before the shock and $\varphi_2$ being the phase after the shock, and the dashed branch - the shocks moving to the left, with the subscripts $1$ and $2$ interchanged.
It follows from the inequality (\ref{sho}) that
on the solid branch the physically relevant part is $-\varphi_1<\varphi_2<\varphi_1$.
Similarly, on the dashed branch the physically relevant parts are $\varphi_2<-\varphi_1$
and $\varphi_2>\varphi_1$.

\section{Riemann method of characteristics}
\label{y}

Multiplying  (\ref{ave9a}) by $\pm\sqrt{\cos\varphi}$ and adding to  (\ref{ave9b})
we obtain
\begin{subequations}
\label{e9}
\begin{alignat}{4}
\frac{\partial Q}{\partial\tau}+\sqrt{\cos\varphi}\frac{\partial \varphi}{\partial\tau}+\sqrt{\cos\varphi}\left(\frac{\partial Q}{\partial Z}
 +\sqrt{\cos\varphi}\frac{\partial \varphi}{\partial Z}\right)&=0.\\
\frac{\partial Q}{\partial\tau}-\sqrt{\cos\varphi}\frac{\partial \varphi}{\partial\tau}-\sqrt{\cos\varphi}\left(\frac{\partial Q}{\partial Z}
 -\sqrt{\cos\varphi}\frac{\partial \varphi}{\partial Z}\right)&=0.
\end{alignat}
\end{subequations}
After we introduce  the new variables
\begin{subequations}
\label{ri}
\begin{alignat}{4}
J_{+}&=\int\sqrt{\cos\varphi}d\varphi+Q,\\
J_{-}&=\int\sqrt{\cos\varphi}d\varphi-Q,
\end{alignat}
\end{subequations}
called the Riemann invariants Eq. (\ref{e9}) takes the form
\begin{subequations}
\label{g8}
\begin{alignat}{4}
\left(\frac{\partial }{\partial\tau}+\sqrt{\cos\varphi}\frac{\partial}{\partial Z}
 \right)J_+&=0,\label{g8a}\\
\left(\frac{\partial }{\partial\tau}-\sqrt{\cos\varphi}\frac{\partial}{\partial Z}
 \right)J_-&=0,\label{g8b}
\end{alignat}
\end{subequations}
where $\varphi$ should be understood as the function of $J_+$ and $J_-$ obtained by
solving (\ref{ri}).
Note that the differential operators acting on $J_+$ and $J_-$ are just the operators of differentiation
along the characteristics
\begin{eqnarray}
C_{+/-}: \frac{dZ}{d\tau}&=\pm\sqrt{\cos\varphi}
\end{eqnarray}
 in the $Z\tau$-plane. Thus we see that $J_+$ and $J_-$ remain
constant along each characteristic $C_+$ or $C_-$ respectively. We can also say that small
perturbations of $J_+$ are propagated only along the characteristics C+, and those of $J_-$
only along $C_-$ \cite{landau2}.

\end{appendix}

\end{document}